\RequirePackage[mathlines]{lineno}
\documentclass[twocolumn,showpacs,aps,prl]{revtex4-2}
\usepackage{overpic,graphicx}
\usepackage{dcolumn}
\usepackage{bm}
\usepackage{rotating}
\usepackage{subfigure}
\usepackage{color}
\usepackage[bookmarksnumbered, pdfstartview=FitH,colorlinks,urlcolor=blue, citecolor=blue,linkcolor=blue,] {hyperref}
\usepackage{lineno}
\usepackage{multirow}
\usepackage{amsmath}
\usepackage{makecell}

\newcommand{\jpsi}{J/\psi}

\begin{document}

\title{\bf \boldmath
First measurement of $\Lambda N$ inelastic scattering with $\Lambda$ from $e^+ e^- \to \jpsi \to \Lambda \bar{\Lambda}$
}

\author{
\begin{small}
\begin{center}
M.~Ablikim$^{1}$, M.~N.~Achasov$^{5,b}$, P.~Adlarson$^{75}$, X.~C.~Ai$^{81}$, R.~Aliberti$^{36}$, A.~Amoroso$^{74A,74C}$, M.~R.~An$^{40}$, Q.~An$^{71,58}$, Y.~Bai$^{57}$, O.~Bakina$^{37}$
% ...
\\
\vspace{0.2cm}
(BESIII Collaboration)\\
\vspace{0.2cm} {\it
$^{1}$ Institute of High Energy Physics, Beijing 100049, People's Republic of China\\
% ...
$^{5}$ Budker Institute of Nuclear Physics SB RAS (BINP), Novosibirsk 630090, Russia\\
% ...
$^{36}$ Johannes Gutenberg University of Mainz, Johann-Joachim-Becher-Weg 45, D-55099 Mainz, Germany\\
$^{37}$ Joint Institute for Nuclear Research, 141980 Dubna, Moscow region, Russia\\
% ...
$^{40}$ Liaoning Normal University, Dalian 116029, People's Republic of China\\
% ...
$^{57}$ Southeast University, Nanjing 211100, People's Republic of China\\
$^{58}$ State Key Laboratory of Particle Detection and Electronics, Beijing 100049, Hefei 230026, People's Republic of China\\
% ...
$^{71}$ University of Science and Technology of China, Hefei 230026, People's Republic of China\\
% ...
$^{74}$ University of Turin and INFN, (A)University of Turin, I-10125, Turin, Italy; (B)University of Eastern Piedmont, I-15121, Alessandria, Italy; (C)INFN, I-10125, Turin, Italy\\
$^{75}$ Uppsala University, Box 516, SE-75120 Uppsala, Sweden\\
% ...
$^{81}$ Zhengzhou University, Zhengzhou 450001, People's Republic of China\\
\vspace{0.2cm}
% ...
$^{b}$ Also at the Novosibirsk State University, Novosibirsk, 630090, Russia\\
% ...
}\end{center}

\vspace{0.4cm}
\end{small}
}

\begin{abstract} 

	Using an $e^+ e^-$ collision data sample of $(10087 \pm 44)\times10^6 ~\jpsi$ events taken at the center-of-mass energy of $3.097~\rm{GeV}$ by the BESIII detector at the BEPCII collider, the process $\Lambda+N \rightarrow \Sigma^+ + X$ is studied for the first time employing a novel method. The $\Sigma^{+}$ hyperons are produced by the collisions of $\Lambda$ hyperons from $J/\psi$ decays with nuclei in the material of the BESIII detector. The total cross section of $\Lambda + ^{9}{\rm Be} \rightarrow \Sigma^+ + X$ is measured to be $\sigma = (37.3 \pm 4.7 \pm 3.5)~{\rm mb}$ at $\Lambda$ beam momenta within $[1.057, 1.091]~{\rm GeV}/c$, where the uncertainties are statistical and systematic, respectively. This analysis is the first  study of $\Lambda$-nucleon interactions at an $e^+ e^-$ collider, providing  information and constraints relevant for the strong-interaction potential, the origin of color confinement, the unified model for baryon-baryon interactions, and the internal structure of neutron stars.

\end{abstract}


\maketitle

\oddsidemargin  -0.2cm
\evensidemargin -0.2cm


Describing baryon-baryon interactions within a unified model has always been a challenge in both particle and nuclear physics~\cite{Vidana:2018bdi,Hiyama:2018lgs,Gal:2016boi,Tolos:2020aln}. Strong constraints and  well-established models exist for nucleon-nucleon interactions~\cite{Vidana:2018bdi,Hiyama:2018lgs}, but there are still difficulties in precisely modeling hyperon-nucleon scattering, especially hyperon-hyperon interactions, due to the lack of experimental measurements. Until now, there have only been a few measurements for hyperon-nucleon scattering~\cite{Eisele:1971mk,Sechi-Zorn:1968mao,Alexander:1968acu,Kadyk:1971tc,Hauptman:1977hr,KEK-PSE-251:1997cno,KEK-PS-E289:2000ytt,Ahn:2005jz,J-PARCE40:2021qxa,J-PARCE40:2021bgw,CLAS:2021gur,J-PARCE40:2022nvq, BESIII:2023clq}, and only one for hyperon-hyperon scattering~\cite{ALICE:2022uso}, leaving theoretical models largely unconstrained~\cite{Haidenbauer:2005zh, Rijken:2010zzb, Polinder:2006zh,Haidenbauer:2013oca,Haidenbauer:2015zqb, Haidenbauer:2018gvg, Haidenbauer:2019boi, Haidenbauer:2023qhf, Li:2016paq, Li:2016mln, Ishii:2006ec,Ishii:2012ssm, Beane:2006gf,Beane:2010em, Schaefer:2005fi, Fujiwara:2006yh}.

The properties of hyperons in dense matter have attracted much interest due to their close connection with hypernuclei and the hyperon component in neutron stars~\cite{Tolos:2020aln}. Hyperons may exist within the inner layer of neutron stars whose structure strongly depends on the equation of state (EOS) of nuclear matter at supersaturation densities~\cite{Lattimer:2000nx}. The appearance of hyperons in the core softens the EOS, resulting in neutron stars with masses lower than 2$M_\odot$~\cite{Lonardoni:2014bwa}, where $M_\odot$ is the mass of the sun. However, studies based on observations from the LIGO and Virgo experiments~\cite{LIGOScientific:2018cki} indicate that the EOS can support neutron stars with masses above $1.97M_\odot$.  This is the so-called ``hyperon puzzle in neutron stars", warranting further experimental and theoretical studies.

The first attempts to measure hyperon-nucleon interactions ($\Lambda p \rightarrow \Lambda p$, $\Sigma^- p \rightarrow \Sigma^- p / \Lambda n / \Sigma^0 n$ and $\Sigma^+ p \rightarrow \Sigma^+ p$) were made during the 1960s and 1970s using hyperons with momenta less than $1~{\rm GeV}/c$~\cite{Eisele:1971mk,Sechi-Zorn:1968mao,Alexander:1968acu,Kadyk:1971tc,Hauptman:1977hr}. After a gap of about 20 years, further studies of elastic and inelastic scatterings between hyperons and nucleons were performed using multiple kinds of hyperons with a variety of beam energies~\cite{KEK-PSE-251:1997cno,KEK-PS-E289:2000ytt,Ahn:2005jz,J-PARCE40:2021qxa,J-PARCE40:2021bgw,CLAS:2021gur,J-PARCE40:2022nvq,ALICE:2022uso}. The uncertainties on these measurements were in general large. On the theoretical side, many models have been proposed to describe the hyperon-nucleon and hyperon-hyperon interactions, including the meson-exchange model (with J\"{u}lich~\cite{Haidenbauer:2005zh} or Nijmegen~\cite{Rijken:2010zzb} potentials), chiral effective field theory ($\chi$EFT) approaches~\cite{Polinder:2006zh,Haidenbauer:2013oca,Haidenbauer:2015zqb, Haidenbauer:2018gvg, Haidenbauer:2019boi, Haidenbauer:2023qhf, Li:2016paq, Li:2016mln}, calculations on the lattice from HALQCD~\cite{Ishii:2006ec,Ishii:2012ssm} and NPLQCD~\cite{Beane:2006gf,Beane:2010em}, low-momentum models~\cite{Schaefer:2005fi} and quark-model approaches~\cite{Fujiwara:2006yh}.

Experimental studies of hyperon-nucleon interactions are challenged by the difficulty of obtaining a stable hyperon beam. Firstly, the lifetime of ground-state hyperons is usually of order ${\cal O}(10^{-10})~{\rm s}$ due to the weak decay, which is too short to provide a stable beam. Meanwhile, hyperons historically used for fixed-target experiments are commonly produced in the collisions between incident protons or $K$ mesons and the target material, with a high background level. Compared with fixed-target experiments, many more hyperons are accessible from the decay of charmonia produced at $e^+ e^-$ colliders, which have rarely been used to study hyperon-nucleon scattering because of the lack of both specialized targets and any practical experimental approach.  Furthermore, the large number of antihyperons produced in pairs with hyperons bring exciting prospects for probing little-studied antihyperon-nucleon interactions. In this work,  the $\Lambda+ N \rightarrow \Sigma^+ + X$ process is studied for the first time by a novel method~\cite{Dai:2022wpg, Yuan:2021yks}, where $N$ denotes a certain kind of nucleus and $X$ refers to any possible particles produced accompanying the $\Sigma^+$, using  $\Lambda \bar{\Lambda}$ pairs from the decay of $(10087\pm44)\times10^{6}$ $\jpsi$ events collected by BESIII~\cite{BESIII:2021cxx, Li:2016tlt}. This method has been applied in a recent BESIII study of $\Xi^0$-nucleus interaction~\cite{BESIII:2023clq}. Thanks to the ``Double Tag" method, a nearly monochromatic hyperon beam of $\Lambda$, $\Sigma^+$, $\Sigma^-$, $\Xi^-$, $\Xi^0$, $\Omega^-$, and their anti-particles from charmonia decay are accessible that allows for the study of hyperon-nucleon interactions at $e^+ e^-$ colliders.

The BESIII detector is a magnetic spectrometer~\cite{BESIII:2009fln} located at the Beijing Electron Positron Collider (BEPCII). The cylindrical core of the BESIII detector consists of a helium-based multilayer drift chamber (MDC), a plastic scintillator time-of-flight system (TOF), and a CsI(Tl) electromagnetic calorimeter (EMC), which are all enclosed in a superconducting solenoidal magnet providing a 1.0~T magnetic field. Before particles produced in $e^+ e^-$ collisions enter the spectrometer, they pass through the beam pipe and the inner wall of the MDC, which constitute the scattering targets in the present study. The target structure and $\Lambda$ trajectory are shown in Fig.~\ref{fig:target}, where the $\Lambda$ can scatter elastically or inelastically with the nuclei inside these objects. The target is made of multiple layers with different materials, with more detailed information given  in Section I of the Supplemental Material~\cite{supplemental}.

\begin{figure}[!htbp]
	\centering
	\includegraphics[width=0.45\textwidth]{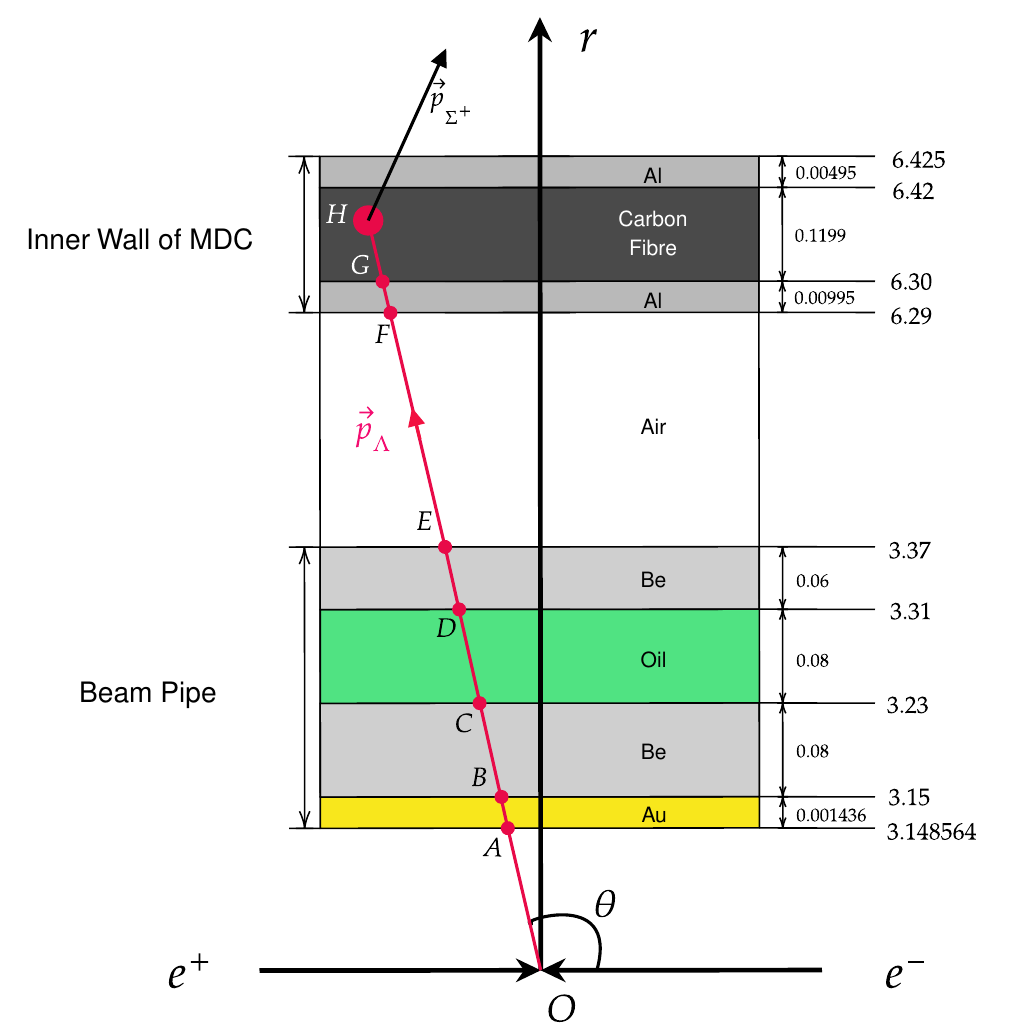}
	\caption{Illustration of target structure and $\Lambda$ trajectory inside the target. The target material, composed of the beam pipe and inner wall of MDC, consists of multiple layers of material, including gold, beryllium, oil, aluminum and carbon fiber. $O$ is the interaction point of $e^+ e^-$ collision. The horizontal axis is the $e^+e^-$ beam line and the vertical axis ($r$ axis) denotes the distance away from the beam line. 
The radius and length of the inner wall of beam pipe from and along the z-axis are 3.148564 $\rm{cm}$ and 100 $\rm{cm}$, respectively~\cite{BESIII:2009fln}.
The position and thickness of each layer are listed in the figure, where the unit is centimeter. $\theta$ is the angle between incident $\Lambda$ and $z$ axis. $H$ is the scattering points of $\Lambda$ and nucleon so that $AB$, $BC$, ..., $GH$ are the $\Lambda$ track lengths in each layer, where the sum is the total track length inside the target.}
	\label{fig:target}
\end{figure}

Using a {\sc geant4}-based~\cite{geant4} Monte Carlo (MC) package, simulated samples are produced incorporating the geometric description~\cite{detvis} of the BESIII detector and the detector response. An inclusive MC sample containing 10 billion $\jpsi$ decays is used to investigate the potential backgrounds. The production of the $\jpsi$ resonance is simulated by the MC event generator {\sc kkmc}~\cite{ref:kkmc}, where the beam-energy spread and initial-state radiation (ISR) in the $e^+ e^-$ annihilation have been taken into account. The known decay modes are generated by {\sc evtgen}~\cite{ref:evtgen, ref:evtgen2} using branching fractions taken from the Particle Data Group (PDG)~\cite{ParticleDataGroup:2022pth}, while the unknown decay modes are modeled with  {\sc lundcharm}~\cite{ref:lundcharm, ref:lundcharm2}. A signal MC sample with one million $\Lambda N \rightarrow \Sigma^{+} X,~\Sigma^+ \rightarrow p \pi^{0}$ events is generated to estimate the detection efficiency. The angular distributions of $\jpsi \rightarrow \Lambda \bar{\Lambda}$ and $\bar{\Lambda} \rightarrow \bar{p} \pi^+$ are described by the recently measured decay parameters of the $\jpsi$ and $\Lambda$ hyperon~\cite{BESIII:2022qax} and $\Lambda N \rightarrow \Sigma^{+} X,~\Sigma^+ \rightarrow p \pi^{0}$ processes are simulated by the Bertini intranuclear cascade model~\cite{Wright:2015xia} of the QGSP\_BERT physics list defined in {\sc geant4}~\cite{geant4}.


Since $\Lambda$ and $\bar{\Lambda}$ are produced in pairs from $\jpsi \to \Lambda \bar{\Lambda}$ decays, the detection of a single $\bar{\Lambda}$ in an event (called ``single-tag") implies that the recoiling system is a monochromatic $\Lambda$. In this analysis, the $\bar{\Lambda}$ hyperon is reconstructed via $\bar{\Lambda} \to \bar{p} \pi^+$ and the yield of single-tagged events is obtained by fitting to the recoil-mass distribution of the $\bar{\Lambda}$, denoted as $N_{\rm ST}$. The recoil mass of the $\bar{\Lambda}$ is defined as $RM_{\bar{p}\pi^+} = \sqrt{|p_{e^+ e^-} - p_{\bar{p}} -p_{\pi^+}|^2}$, where $p_{e^+ e^-}$, $p_{\bar{p}}$ and $p_{\pi^+}$ refer to the four-momenta of the initial $e^+ e^-$, $\bar{p}$ and $\pi^+$ particles, respectively. The recoiling $\Lambda$ produced together with the reconstructed $\bar{\Lambda}$ can scatter inelastically with the nucleons in the material and produce a $\Sigma^+$ hyperon. We search for such particles through the decay $\Sigma^+ \to p \pi^0$ among the other tracks and showers in the event, excluding those used to reconstruct the single-tagged $\bar{\Lambda}$.  The number of double-tagged events ($N_{\rm DT}$) containing both a reconstructed  $\bar{\Lambda}$ and $\Sigma^+$ is given by
\begin{equation}
	\label{equ:DT_yield}
		N_{\rm DT} = \mathcal{L}_{\Lambda} \cdot \sigma(\Lambda N \to \Sigma^+ X) \cdot \mathcal{B}(\Sigma^+ \to p \pi^0) \cdot \epsilon_{\rm sig},
\end{equation}
where $\sigma(\Lambda N \to \Sigma^+ X)$ is the cross section of the inelastic process $\Lambda N \to \Sigma^+ X$, $\mathcal{B}(\Sigma^+ \to p \pi^0)$ is the branching fraction of $\Sigma^+ \to p \pi^0$ decay and $\epsilon_{\rm sig}$ denotes the efficiency of the double-tag reconstruction.  The ``effective luminosity'' $\mathcal{L}_{\Lambda}$ is a specially defined quantity to describe the property of the target and the behavior of the incident $\Lambda$ particle inside the target, which is influenced by several other parameters~\cite{CLAS:2021gur} . Considering the target composition shown in Fig.~\ref{fig:target}, $\mathcal{L}_{\Lambda}$ is calculated event by event as
\begin{equation}
        \label{equ:cal_eff_lum}
	\mathcal{L}_{\Lambda} = N_{\rm ST} \cdot \frac{N_{A}}{N_{\rm ST}^{\rm MC}} \cdot \sum_{j}^{7} \sum_{i}^{N_{\rm ST}^{\rm MC}} \frac{\rho_{T}^{j} \cdot l^{ij}}{M^{j}} \cdot \mathcal{R}_{\sigma}^{j},
\end{equation}
where $N_{A}$ is Avogadro's number~\cite{Millikan:1913zz}, $N_{\rm ST}^{\rm MC}$ is the total number of single-tagged events in the signal MC sample, $l^{ij}$ is the path length of the incident $\Lambda$ of the $i_{\rm th}$ event inside the $j_{\rm th}$ layer, and $M^j$ and $\rho_{T}^j$ are the molar mass~\cite{NIST:2023} and density, respectively, of the $j_{\rm th}$ layer. Since the contribution from each layer and the cross section for different kinds of nuclei are not the same, the ratio of the scattering cross section (${\mathcal R}_{\sigma}$) of incident $\Lambda$ for each category of materials is necessary for normalization.
In this Letter, we present the measured cross section of the $\Lambda + ^{9}{\rm Be} \rightarrow \Sigma^+ + X$ reaction, with the cross section of each material normalized to that of beryllium. The choice of beryllium as the normalization reference is due to its common use as the target in fixed-target experiments and its significance as the main material of the beam pipe in the BESIII.
In the case of low and intermediate energy it is assumed that inelastic scattering occurs with single protons on the surface of the nucleus~\cite{Barton:1982dg, Cooper:1995ix, WA89:1997dmp, Botta:2001fu, Astrua:2002zg, Lee:2018epd}. Then ${\mathcal R}_{\sigma}$ is proportional to $A^{\frac{2}{3}} \times \frac{Z}{A} = \frac{Z}{A^{\frac{1}{3}}}$, where $A$ and $Z$ are the numbers of nucleons and protons in a single nucleus, respectively. If the material of a certain layer contains multiple kinds of nuclei, ${\mathcal R}_{\sigma}$ must be weighted by the ratio of the numbers of different nuclei per unit volume of the material. The detailed derivation and calculation of $\mathcal{L}_{\Lambda}$ can be found in Section II in Supplemental Material~\cite{supplemental}.

According to Eq.~(\ref{equ:DT_yield}), the cross section for interaction with the Be nucleus can be determined as 
\begin{equation}
	\label{equ:cross_section}
	\sigma({\rm Be}) = \frac{N_{\rm DT}}{\epsilon_{\rm sig} \cdot {\mathcal{L}_{\Lambda}}} \cdot \frac{1}{{\mathcal B}(\Sigma^+ \to p \pi^0)}.
\end{equation}

Since we only reconstruct the $\Sigma^+$ on the double-tag side, there may also be contributions from the interactions between the $\Lambda$ and neutrons. However, the contributions can only arise from  three-body reactions such as  $\Lambda n \to \Sigma^{+} n \pi^{-}$.
The total energy in the center-of-mass frame of $\Lambda$ and a stationary neutron is [2.240, 2.249] GeV, while the lowest total energy for the final state $\Sigma^{+}n\pi^-$ is 2.269 GeV. Therefore, this process can only occur when the neutron have relatively large Fermi momenta. As a result, the cross-section is suppressed due to the limited phase space.
Similar reactions such as $\Lambda p \to \Sigma^- p \pi^+/\Sigma^+ p \pi^-$ have been studied~\cite{Hauptman:1977hr}, which for $\Lambda$ momentum within $[1.057, 1.091]~{\rm GeV}/c$ are at least one magnitude smaller than the measured cross section of $\Lambda p \to \Sigma^+ X$.  Therefore, we neglect the contribution from  $\Lambda n$ reactions.


We now describe the selection of signal events. Charged tracks detected in the MDC are required to have a polar angle ($\theta$) satisfying $\vert\!\cos\theta\vert<0.93$ with respect to the positron beam. The number of good charged tracks must be at least two.

\begin{figure}[!htbp]
	\centering
	\includegraphics[width=0.45\textwidth]{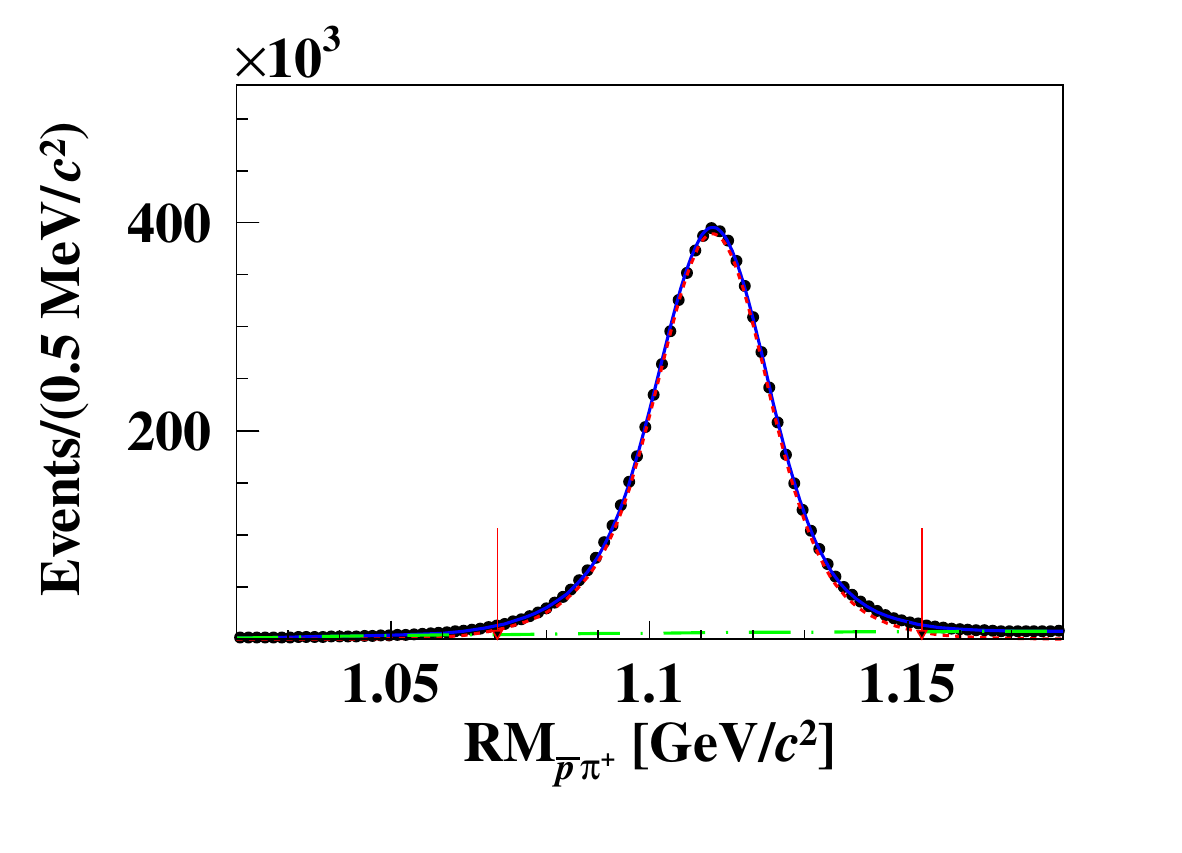}
	\caption{The  $RM_{\bar{p} \pi^+}$ distribution with the fit result superimposed. The black dots with error bars represent the data. The blue solid line is the total fit. The dashed red line is the signal of the single-tag selection and the dot-dashed green line is the background. The red arrows indicate the signal range.   }
	\label{fig:ST_yield}
\end{figure}

For the single-tag side, $\bar{\Lambda}$ is reconstructed from its decay to $\bar{p}$ and $\pi^+$, which are identified using the measured information in the MDC and TOF. The combined likelihoods ($\mathcal{L}$) under the proton (antiproton), pion and kaon hypotheses are calculated. The $\bar{p}$ candidates are required to satisfy $\mathcal{L}(p) > \mathcal{L}(K)$ and $\mathcal{L}(p) > \mathcal{L}(\pi)$, whereas the $\pi^+$ candidates are required to satisfy $\mathcal{L}(\pi) > \mathcal{L}(K)$. A vertex fit is performed to constrain all possible $\bar{p}\pi^+$ combinations to a common vertex. The decay length of $\bar{\Lambda}$ is calculated as the distance between the fitted vertex and the interaction point of the $e^+ e^-$ collision (IP). The $\bar{p} \pi^+$ combinations with a vertex-fit $\chi^2$ lower than 200 and a decay length larger than 0 are regarded as $\bar{\Lambda}$ candidates. To further suppress background, the invariant masses of the $\bar{p} \pi^+$ combinations are required to lie within $[1.111, 1.120]~{\rm GeV}/c^2$ and  $RM_{\bar{p} \pi^+}$ is required to be within $[1.071, 1.153]~{\rm GeV}/c^2$. If there are multiple candidates passing all the selection criteria above, we select the $\bar{p} \pi^+$ combination with the minimum vertex-fit $\chi^2$  as the best candidate. The number of single-tagged events is determined to be  $7207565\pm3741$  by fitting to the  $RM_{\bar{p} \pi^+}$ distribution using the sum of two Gaussian distributions and a second-order Chebyshev polynomial,  as shown in Fig.~\ref{fig:ST_yield}. The efficiency of the single-tag selection is $(52.16 \pm 0.10)\%$.

In the double-tag selection $\Sigma^+$ candidates are reconstructed in the $p \pi^0$ final state. The $p$ is selected using the same criteria as the single-tag selection and the $\pi^0$ is reconstructed through  EMC showers. The deposited energy of each shower must be more than $25~{\rm MeV}$ in the barrel region ($\vert\!\cos\theta\vert<0.80$) and more than $50~{\rm MeV}$ in the end-cap region ($\vert\!\cos\theta\vert<0.92$) of the EMC. To suppress electronic noise and showers unrelated to the event, the difference between the EMC time and the event-start time is required to be within $[0, 700]~{\rm ns}$. The angle between photons and all the other charged tracks should be larger than $10^{\circ}$ to suppress the photons from the radiation of charged tracks and other processes. Then a kinematic fit is performed to all possible combinations of two showers by constraining the invariant mass $M_{\gamma \gamma}$ to  the known $\pi^0$ mass~\cite{ParticleDataGroup:2022pth}. The combination with the minimum $\chi^2$ from this fit is chosen as the $\pi^0$ candidate. The $\Sigma^+$ candidate is selected from  $p \pi^0$ combinations with invariant mass $M_{p \pi^0} \in [1.12, 1.25]~{\rm GeV}/c^2$  that has the  maximum value of $\mathcal{L}(p)$ for the proton candidate.

If the incident $\Lambda$ does not scatter with any nucleons, the recoil mass of $\bar{p} \pi^+ p$, $RM_{\bar{p} \pi^+ p}$, should be around the known mass of $\pi^-$~\cite{ParticleDataGroup:2022pth} according to the kinematic constraint. 
For inelastic scattering events, $RM_{\bar{p} \pi^+ p}$ has a tendency to be negative, as seen with the signal MC sample and discussed in Section III of the Supplemental Material~\cite{supplemental}.  This behavior is explained by the additional mass and momentum contributed by nucleons in the reaction. To further suppress  background, $RM_{\bar{p} \pi^+ p}$ is required to be negative. In addition, the normal event of $J/\psi \to \Lambda \bar{\Lambda}$ without the hyperon-nucleus scattering can be rejected by identifying a $\Lambda$ in the double-tag side after the $\Sigma^+$ candidate reconstruction using all possible $p \pi^-$ combinations. The $\Lambda$ candidate is rejected if $M_{p \pi^-} \in [1.108, 1.124]~{\rm GeV}/c^2$.

\begin{figure}[!htbp]
	\centering
	\includegraphics[width=0.45\textwidth]{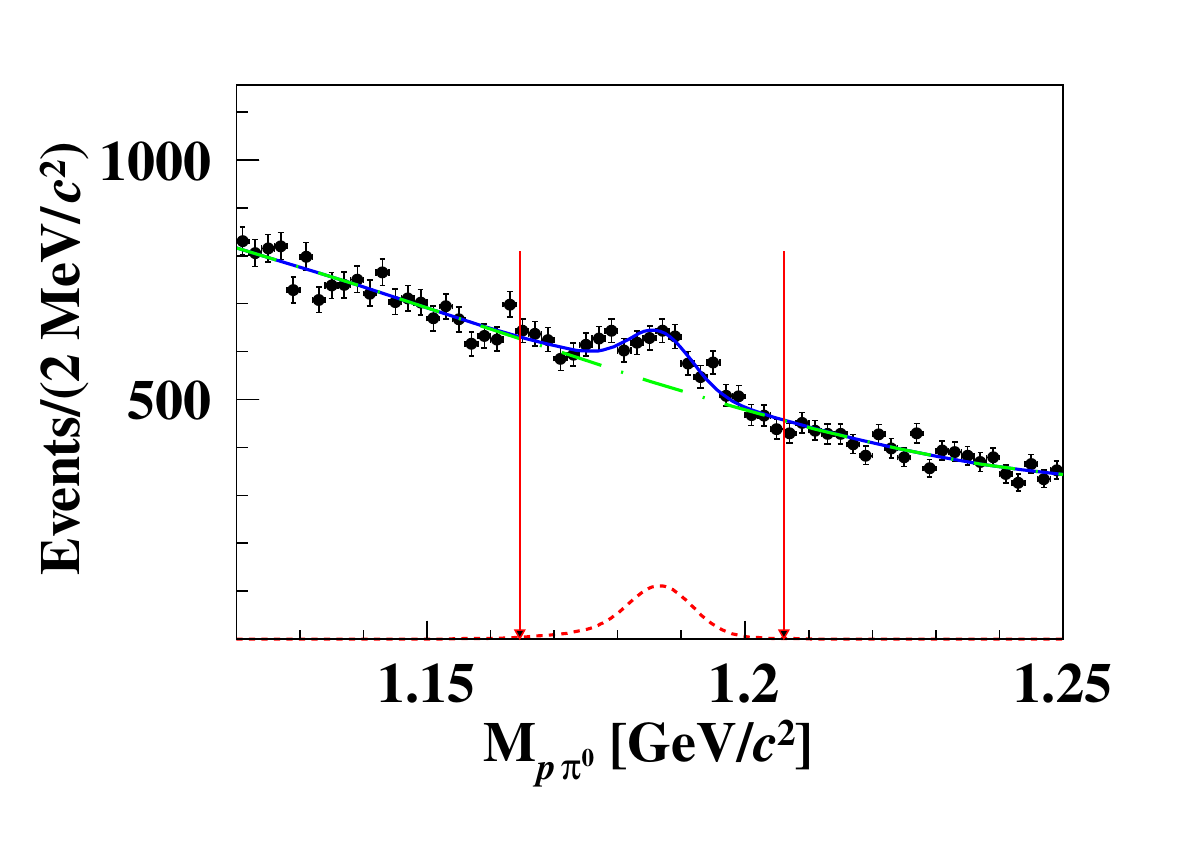}
	\caption{The $M_{p \pi^0}$ distribution with the fit result overlaid. The black dots with error bars represent the data. The blue solid line is the total fit. The dashed red line is the signal and the dot-dashed green line is the background. The red arrows indicate the signal range.}
	\label{fig:DT_yield}
\end{figure}

The number of double-tagged events is found to be $795\pm101$ by fitting to the distribution of $M_{p \pi^0}$ using a sum of two Gaussian distributions and a third-order Chebyshev polynomial, as is shown in Fig.~\ref{fig:DT_yield}. The parameters and fraction of the two Gaussian distributions are fixed to those obtained from the same fit to the signal MC. The efficiency of the double-tag selection is estimated to be $\epsilon_{\rm sig}=(24.32 \pm 0.15)\%$ by fitting the signal MC sample.

The cross section of $\Lambda + ^{9}{\rm Be} \rightarrow \Sigma^+ + X$ is determined through Eq.~(\ref{equ:cross_section}) to be $\sigma(\Lambda + ^{9}{\rm Be} \rightarrow \Sigma^+ + X) = (37.3 \pm 4.7_{\rm stat.} \pm 3.5_{\rm syst.})~{\rm mb}$, where ${\mathcal B}(\Sigma^+ \to p \pi^0)$ is taken from PDG~\cite{ParticleDataGroup:2022pth}. Table~\ref{tab:paras} lists the inputs used in the calculation.

\begin{table}[htbp]
	\centering
	\caption{Inputs used to calculate the cross section of $\Lambda + ^{9}{\rm Be} \rightarrow \Sigma^+ + X$.}
	\begin{tabular}{p{4cm}<{\centering}p{4cm}<{\centering}}
		\hline
		\hline
		Parameter		&		Value		\\
		\hline
		$N_{\rm DT}$		&		$795\pm101$		\\
		$\epsilon_{\rm sig}$	&		24.32\%			\\
		$\mathcal{L}_{\Lambda}$	&		$(17.00\pm0.01)\times10^{28}~{\rm cm^{-2}}$	\\
		$\mathcal{B}(\Sigma^+ \to p \pi^0)$	&	$(51.57\pm0.30)\%$	\\
		\hline
		\hline
	\end{tabular}
	\label{tab:paras}
\end{table}



The systematic uncertainty for the measured cross section is associated with the knowledge of the tracking and PID efficiencies of charged particles, the reconstruction efficiency of $\pi^0$ mesons, the number of single-tagged events, the efficiency of the requirement of $RM_{\bar{p} \pi^+ p}$ and $M_{p \pi^-}$,the  angular distribution of $\jpsi \to \Lambda \bar{\Lambda}$ and $\Sigma^+$, the size of the signal MC sample, the measured interaction point of the $e^+ e^-$ collision, the method to fit $M_{p \pi^0}$ distribution and the luminosity ($\mathcal{L}_{\Lambda}$) estimation. The assumption concerning the ratio of cross sections for different kinds of nucleus also introduces a systematic uncertainty.

The systematic uncertainty related to the tracking and PID efficiencies of proton are both 1.0\%~\cite{BESIII:2021emv}, and for $\pi^0$ reconstruction is 1.0\%~\cite{BESIII:2010ank}. The systematic uncertainty associated with the number of the single-tagged events is determined to be 0.8\% from the inclusive MC. The systematic uncertainties from the $RM_{\bar{p} \pi^+ p}$ and $M_{p \pi^-}$ requirements are tested by varying the criteria around the baseline settings to re-obtain the measured cross section. The changes of the cross section are denoted as $\Delta = |\sigma - \sigma_{\rm sys.}|$, where $\sigma$ and $\sigma_{\rm sys.}$ refer to the baseline results and the results after changing the criteria. Also calculated are the uncorrelated uncertainties $\omega_{\rm uc.} = \sqrt{|\omega_{\sigma}^{2}-\omega_{\sigma, {\rm sys.}}^{2}|}$, where $\omega_{\sigma}$ and $\omega_{\sigma, {\rm sys.}}$ correspond to the fit uncertainties of the baseline and systematic test results, respectively. Since the ratio $\Delta/\omega_{\rm uc.}$ does not show a trending behavior and is less than two, these two possible sources of systematic bias are considered to be negligible~\cite{Barlow:2002yb}. The systematic uncertainty associated with the knowledge of  the angular distribution of $\jpsi \to \Lambda \bar{\Lambda}$ production is estimated by varying the decay parameters $\alpha_{\jpsi}$, $\Delta \Phi$, $\alpha_{\Lambda}$ and $\alpha_{\bar{\Lambda}}$ within one standard deviation~\cite{BESIII:2022qax} and generating new MC data sets to calculate the cross section. The systematic uncertainty from this source can be ignored compared with the statistical uncertainty. The systematic uncertainty associated with the angular distribution of $\Sigma^+$ baryons is evaluated by reweighting the angular distribution of the signal MC to that of the data and measuring the cross section again, which is determined to be 1.3\% as the difference between the re-obtained cross section and the baseline value.

The double-tag efficiency has a systematic uncertainty of 0.6\% arising from the fitted number of double-tagged events in the signal MC sample. The interaction point (IP) of the $e^+ e^-$ collision is used to calculate the track length of $\Lambda$ inside the target. According the measured result, the interaction point is distributed around the coordinate origin with an uncertainty of 0.2~cm. We change IP within $\pm 0.2~{\rm cm}$ away from the original position and take the maximum change of the measured cross section as the systematic uncertainty, which is 4.6\%. The systematic uncertainty from the fit method of $M_{p \pi^0}$ distribution is estimated by changing the background shape from a third-order Chebyshev polynomial to fourth- and fifth-order ones, and assign the uncertainty to  be 3.6\% as the maximum difference from the baseline result. As well as  scattering  inside the beam pipe and the inner wall of the  MDC, the $\Lambda$ may also scatter with a nucleus inside the cooling devices of the BESIII spectrometer. When calculating the luminosity $\mathcal{L}_{\Lambda}$, the cooling pipe is not considered due to its much more complex structure and having less material than the beam pipe and inner wall of MDC. We assume a conservative model of the cooling devices and measure the cross section again. The difference from the baseline result is assigned as the systematic uncertainty, which is 6.1\%.

In order to estimated  the systematic uncertainty caused by the assumption of the ratio of the cross sections for different kinds of nuclei, we measure the cross section again under another assumption that the cross section is proportional to the total number of protons in a nucleus. The difference from our baseline result is taken as the systematic uncertainty, which is determined to be 3.6\%.

The total systematic uncertainty on the measured cross section of $\Lambda + ^{9}{\rm Be} \rightarrow \Sigma^+ + X$ is computed to be 9.5\% by adding the systematic uncertainties listed above in quadrature.


In summary, the inelastic scattering $\Lambda + ^{9}{\rm Be} \rightarrow \Sigma^+ + X$ is studied at BESIII using a novel method. The cross section is measured to be $\sigma(\Lambda + ^{9}{\rm Be} \rightarrow \Sigma^+ + X) = (37.3 \pm 4.7_{\rm stat.} \pm 3.5_{\rm syst.})~{\rm mb}$ for a Be nucleus struck by a $\Lambda$ with momentum within $[1.057, 1.091]~{\rm GeV}/c$. 
Taking 1.93 as the ratio of the cross section of $\Lambda + ^{9}{\rm Be} \rightarrow \Sigma^+ + X$ and $\Lambda + p \rightarrow \Sigma^+ + X$
by assuming the the signal process as a surface reaction~\cite{Barton:1982dg, Cooper:1995ix, WA89:1997dmp, Botta:2001fu, Astrua:2002zg, Lee:2018epd}, the cross section of $\Lambda + p \rightarrow \Sigma^+ + X$ is determined to be $\sigma(\Lambda + p \rightarrow \Sigma^+ + X) = (19.3 \pm 2.4_{\rm stat.} \pm 1.8_{\rm syst.})~{\rm mb}$.

This is the first discovery and cross section measurement of $\Lambda + p \rightarrow \Sigma^+ + n$. By virtue of charge independence, the cross-sections of $\Lambda + p \rightarrow \Sigma^+ + n$ are just twice that of $\Lambda + p \rightarrow \Sigma^0 + p$~\cite{Kadyk:1971tc}. Our results are consistent with previous experiments regarding the cross section measurement of $\Lambda + p \rightarrow \Sigma^0 + p$~\cite{Hauptman:1977hr}.
Additionally, this study represents the first attempt to investigate $\Lambda$-nucleus interactions at an $e^{+} e^{-}$ collider.
The result will be valuable for improving the understanding  of the potential of strong interaction and the origin of color confinement, as well as providing important constraints for the unified model for baryon-baryon interactions.
At BESIII, it is possible to measure the differential cross sections with respect to the momentum of incident hyperons using three-body decays of charmonia with at least one hyperon.

In the future, the Super $\tau$-Charm Facility~\cite{Achasov:2023gey} will produce a $\jpsi$ data set about 100 times larger than the sample collected by BESIII, which will allow for more detailed studies of  the mechanism of hyperon-nucleon interactions.

\section*{\boldmath ACKNOWLEDGMENTS}

The BESIII Collaboration thanks the staff of BEPCII and the IHEP computing center for their strong support. This work is supported in part by National Key R\&D Program of China under Contracts Nos. 2020YFA0406300, 2020YFA0406400; National Natural Science Foundation of China (NSFC) under Contracts Nos. 11635010, 11735014, 11835012, 11935015, 11935016, 11935018, 11961141012, 12022510, 12025502, 12035009, 12035013, 12061131003, 12165022, 12192260, 12192261, 12192262, 12192263, 12192264, 12192265, 12221005, 12225509, 12235017; the Chinese Academy of Sciences (CAS) Large-Scale Scientific Facility Program; the CAS Center for Excellence in Particle Physics (CCEPP); Joint Large-Scale Scientific Facility Funds of the NSFC and CAS under Contract No. U1832207; CAS Key Research Program of Frontier Sciences under Contracts Nos. QYZDJ-SSW-SLH003, QYZDJ-SSW-SLH040; 100 Talents Program of CAS; The Institute of Nuclear and Particle Physics (INPAC) and Shanghai Key Laboratory for Particle Physics and Cosmology; Yunnan Fundamental Research Project under Contract No. 202301AT070162; European Union's Horizon 2020 research and innovation programme under Marie Sklodowska-Curie grant agreement under Contract No. 894790; German Research Foundation DFG under Contracts Nos. 455635585, Collaborative Research Center CRC 1044, FOR5327, GRK 2149; Istituto Nazionale di Fisica Nucleare, Italy; Ministry of Development of Turkey under Contract No. DPT2006K-120470; National Research Foundation of Korea under Contract No. NRF-2022R1A2C1092335; National Science and Technology fund of Mongolia; National Science Research and Innovation Fund (NSRF) via the Program Management Unit for Human Resources \& Institutional Development, Research and Innovation of Thailand under Contract No. B16F640076; Polish National Science Centre under Contract No. 2019/35/O/ST2/02907; The Swedish Research Council; U. S. Department of Energy under Contract No. DE-FG02-05ER41374.

\clearpage

\onecolumngrid
\section{\bf \boldmath
Supplemental material for "First measurement of $\Lambda N$ inelastic scattering with $\Lambda$ from $e^+ e^- \to \jpsi \to \Lambda \bar{\Lambda}$"}
\twocolumngrid

\oddsidemargin  -0.2cm
\evensidemargin -0.2cm

\section{Structure of the target (beam pipe and the inner wall of MDC of BESIII detector)}

The structure of the target used in this analysis is presented in Fig.~\ref{fig:target}. The target consists of multiple layers composed of gold ($^{197}{\rm Au}$), beryllium ($^{9}{\rm Be}$), oil ($m_{^{12}{\rm C}} : m_{^{1}{\rm H}} = 84.923\% : 15.077\%$), aluminum ($^{27}{\rm Al}$) and carbon fiber ($m_{^{12}{\rm C}} : m_{^{1}{\rm H}} : m_{^{16}{\rm O}} = 69.7\% : 0.61\% : 29.69\%$), where $m_{\rm N}$ is the mass fraction of nucleus $N$. In Fig.~\ref{fig:target}, $O$ is the interaction point of the $e^+ e^-$ collision. The vertical axis ($z$ axis) is the direction of $e^+$ beam and the horizontal axis ($r$ axis) denotes the distance away from the $z$ axis. The quantities $r_{i}$ and $t_{i}$ refer to the radius and the thickness of each layer, respectively, $\theta$ is the angle between the incident $\Lambda$ and the $z$ axis, while $H$ is the scattering point of the $\Lambda$ and nucleon so that $AB$, $BC$, ..., $GH$ are the track lengths of the $\Lambda$ in each layer, of which the sum is the total track length inside the target. The radius and the thickness of each layer are listed in Eq.~(\ref{equ:thickness}). Expression~(\ref{equ:density}) gives the density and molar mass of each layer, where $\rho_{T}$ and $M$ refer to density and molar mass, respectively. The molar mass of the oil and carbon fiber in the target is calculated by averaging the molar masses of C, H and O nuclei weighted by the number of each kind of nucleus inside a unit volume.

\begin{equation}
	\label{equ:thickness}
        r_{i}\left\{{
                \begin{array}{l}
                        r_1 = 3.148564~{\rm cm} \\
                        r_2 = 3.15~{\rm cm}     \\
                        r_3 = 3.23~{\rm cm}     \\
                        r_4 = 3.31~{\rm cm}     \\
                        r_5 = 3.37~{\rm cm}     \\
                        r_6 = 6.29~{\rm cm}     \\
                        r_7 = 6.30~{\rm cm}     \\
                        r_8 = 6.42~{\rm cm}     \\
                        r_9 = 6.425~{\rm cm}
                \end{array}}
                \right.
                t_{i}\left\{{
                        \begin{array}{l}
                                t_1 = 0.001436~{\rm cm} \\
                                t_2 = 0.08~{\rm cm}     \\
                                t_3 = 0.08~{\rm cm}     \\
                                t_4 = 0.06~{\rm cm}     \\
                                t_5 = 0.00995~{\rm cm}  \\
                                t_6 = 0.1199~{\rm cm}   \\
                                t_7 = 0.00495~{\rm cm}  \\
                        \end{array}}
                        \right.
\end{equation}

\begin{widetext}
	\begin{equation}
		\label{equ:density}
	        \rho_T,~M\left\{{
	                \begin{array}{lll}
	                        \rho({\rm Au}) = 19.32~{\rm g \cdot cm^{-3}},   &       M({\rm Au}) = 197~{\rm g \cdot mol^{-1}},       &       r_1 \le r \le r_2       \\
	                        \rho({\rm Be}) = 1.848~{\rm g \cdot cm^{-3}},   &       M({\rm Be}) = 9~{\rm g \cdot mol^{-1}},       &       r_2 \le r \le r_3        \\
	                        \rho({\rm Oil}) = 0.81~{\rm g \cdot cm^{-3}},   &       M({\rm Oil}) = 4.51~{\rm g \cdot mol^{-1}},       &       r_3 \le r \le r_4        \\
	                        \rho({\rm Be}) = 1.848~{\rm g \cdot cm^{-3}},   &       M({\rm Be}) = 9~{\rm g \cdot mol^{-1}},       &       r_4 \le r \le r_5        \\
	                        \rho({\rm Al}) = 2.7~{\rm g \cdot cm^{-3}},   &       M({\rm Al}) = 27~{\rm g \cdot mol^{-1}},       &       r_6 \le r \le r_7        \\
	                        \rho({\rm Carb}) = 1.57~{\rm g \cdot cm^{-3}},   &       M({\rm Carb}) = 12.09~{\rm g \cdot mol^{-1}},       &       r_7 \le r \le r_8        \\
	                        \rho({\rm Al}) = 2.7~{\rm g \cdot cm^{-3}},   &       M({\rm Al}) = 27~{\rm g \cdot mol^{-1}},       &       r_8 \le r \le r_9        \\
	                \end{array}}
	                \right.
	\end{equation}
\end{widetext}

\begin{figure}[!htbp]
	\centering
	\includegraphics[bb=0 0 496.000031 307.5, width=0.5\textwidth]{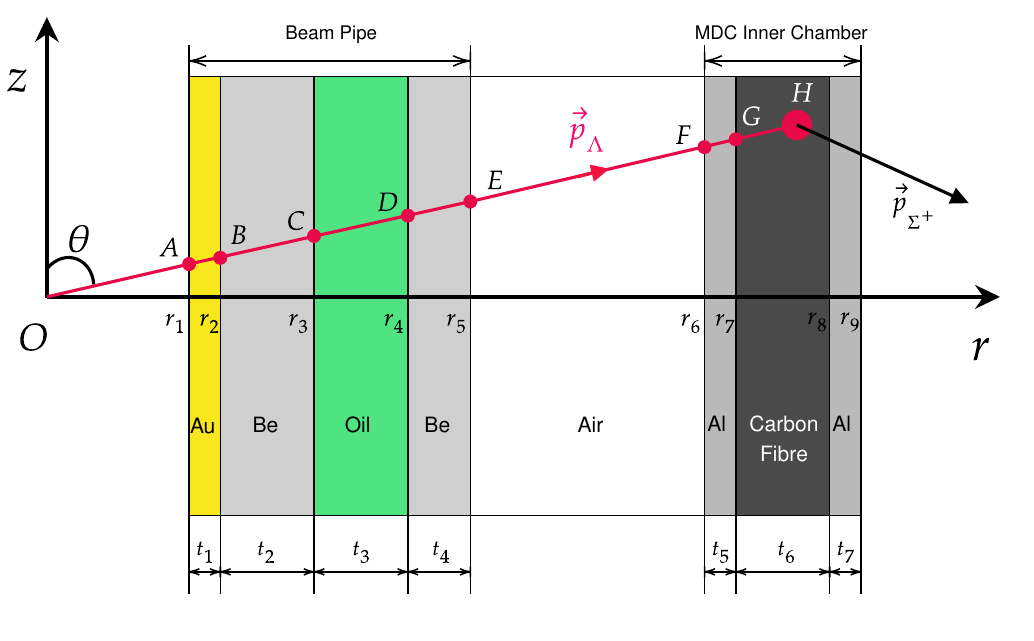}
	\caption{Schematic diagram of the target.}
	\label{fig:target}
\end{figure}

\section{Calculation of $\mathcal{L}_{\Lambda}$}

The formula used to calculate $\mathcal{L}_{\Lambda}$, as firstly introduced by the CLAS collaboration~\cite{CLAS:2021gur_2}, which describes the property of the target and the behavior of the incident $\Lambda$ inside the target, is
\begin{equation}
	\label{equ:effective_lum}
	\mathcal{L}_{\Lambda} = N_{\rm ST} \cdot \frac{N_{A} \cdot \rho_{T} \cdot l}{M},
\end{equation}
where $N_{\rm ST}$ is the number of single-tagged events, $N_{A}$ is the Avogadro constant, $\rho_{T}$ is density of the target, $l$ is the average path length of the $\Lambda$ beam inside the target and $M$ is the molar mass of the target. In this analysis, the target is composed of several layers of different materials. The total value of $\mathcal{L}_{\Lambda}$ is the sum of the contributions of each layer, which gives
\begin{equation}
	\label{equ:total_lum}
	\mathcal{L}_{\Lambda} = \sum_{j}^{7} \mathcal{L}_{\Lambda}^{j} = N_{\rm ST} \cdot N_{A} \cdot \sum_{j}^{7} \frac{\rho_{T}^{j} \cdot l^{j}}{M^{j}} \cdot \mathcal{R}_{\sigma}^{j},
\end{equation}
where $j$ is the index of the layers. Here, $\mathcal{R}_{\sigma}^{j}$ is the factor to normalize the cross sections of $\Lambda N \to \Sigma^+ X$ for different layers into a single layer. In this analysis, we report the cross section for Be nucleus so that $\mathcal{R}_{\sigma}^{j}$ will be unity for Be and model-dependent factors for other kinds of nucleus. According to previous studies,  in the case of low and intermediate energy the inelastic scattering happens mostly with the single protons on the surface of the nucleus ~\cite{Barton:1982dg_2, Cooper:1995ix_2, WA89:1997dmp_2, Botta:2001fu_2, Astrua:2002zg_2, Lee:2018epd_2}.  For this reason, $\mathcal{R}_{\sigma}$ is proportional to $A^{\frac{2}{3}} \times \frac{Z}{A} = \frac{Z}{A^{\frac{1}{3}}}$, where $A$ and $Z$ are the number of nucleons and protons in a single nucleus. Therefore, $\mathcal{R}_{\sigma}$ for each layer is
\begin{equation}
        \label{equ:assum_2}
	{\mathcal R}_{\sigma}^{j}\left\{{
                \begin{array}{l}
                        {\mathcal R}_{\sigma}^{1} = 7.06,         \\
                        {\mathcal R}_{\sigma}^{2} = 1.0,          \\
                        {\mathcal R}_{\sigma}^{3} = 0.789,        \\
			{\mathcal R}_{\sigma}^{4} = 1.0,          \\
                        {\mathcal R}_{\sigma}^{5} = 2.253,        \\
                        {\mathcal R}_{\sigma}^{6} = 1.365,        \\
                        {\mathcal R}_{\sigma}^{7} = 2.253.       \\
                \end{array}}
                \right.
\end{equation}
The values of $\mathcal{R}_{\sigma}$ for oil and carbon fiber in the target are calculated by averaging $\mathcal{R}_{\sigma}$ of C, H and O nuclei weighted by the number of each kind of nucleus inside the unit volume.

The average path length $l^{j}$ inside each layer is calculated using a MC sample including 1 million events of $\jpsi \to \Lambda \bar{\Lambda},~\Lambda N \to \Sigma^+ X,~\bar{\Lambda} \to \bar{p} \pi^+~\Sigma^+ \to p \pi^0$, in which the decay of the $\Lambda$ beam is taken in to consideration. The average path length  is 
\begin{equation}
	\label{equ:average_path_length}
	l^{j} = \frac{\sum_{i}^{N_{\rm ST}^{\rm MC}} l^{ij}}{N_{\rm ST}^{\rm MC}},
\end{equation}
where $N_{\rm ST}^{\rm MC} = 529875$ is the total number of single-tagged events in the MC sample and $l^{ij}$ refers to the path length of the $\Lambda$ of $i_{\rm th}$ event inside the $j_{\rm th}$ layer, which is set to zero if the $\Lambda$ hyperon does not enter the layer. The average path length of each layer is given in Eq.~(\ref{equ:average_length}).
\begin{equation}
	\label{equ:average_length}
        l^{j}\left\{{
                \begin{array}{l}
                        l^{1} = 0.10 \times 10^{-3} ~{\rm cm},         \\
                        l^{2} = 5.76 \times 10^{-3} ~{\rm cm},          \\
                        l^{3} = 5.68 \times 10^{-3} ~{\rm cm},        \\
                        l^{4} = 4.21 \times 10^{-3} ~{\rm cm},          \\
                        l^{5} = 0.43 \times 10^{-3} ~{\rm cm},        \\
                        l^{6} = 5.13 \times 10^{-3} ~{\rm cm},        \\
                        l^{7} = 0.21 \times 10^{-3} ~{\rm cm}.        \\
                \end{array}}
                \right.
\end{equation}

Combining Eq.~(\ref{equ:total_lum}) and Eq.~(\ref{equ:average_path_length}), gives
\begin{equation}
	\label{equ:final_lum}
	\mathcal{L}_{\Lambda} = N_{\rm ST} \cdot \frac{N_{A}}{N_{\rm ST}^{\rm MC}} \cdot \sum_{j}^{7} \sum_{i}^{N_{\rm ST}^{\rm MC}} \frac{\rho_{T}^{j} \cdot l^{ij}}{M^{j}} \cdot \mathcal{R}_{\sigma}^{j}.
\end{equation}

We obtain the values of $\mathcal{L}_{\Lambda}$ for each layer as given in Eq.~(\ref{equ:lum_layer}) with $N_{\rm ST} = 7207565\pm3741$. For the whole target, $\mathcal{L}_{\Lambda} = (17.00\pm0.01) \times 10^{28} ~{\rm cm^{-2}}$, where the uncertainty is propagated from that of $N_{\rm ST}$.
\begin{equation}
        \label{equ:lum_layer}
        {\mathcal L}_{\Lambda}^{j}\left\{{
                \begin{array}{l}
			\mathcal{L}_{\Lambda}^{1} = 0.31 \times 10^{28} ~{\rm cm^{-2}},         \\
                        \mathcal{L}_{\Lambda}^{2} = 5.13 \times 10^{28} ~{\rm cm^{-2}},          \\
                        \mathcal{L}_{\Lambda}^{3} = 3.50 \times 10^{28} ~{\rm cm^{-2}},        \\
                        \mathcal{L}_{\Lambda}^{4} = 3.76 \times 10^{28} ~{\rm cm^{-2}},          \\
                        \mathcal{L}_{\Lambda}^{5} = 0.42 \times 10^{28} ~{\rm cm^{-2}},        \\
                        \mathcal{L}_{\Lambda}^{6} = 3.69 \times 10^{28} ~{\rm cm^{-2}},        \\
                        \mathcal{L}_{\Lambda}^{7} = 0.20 \times 10^{28} ~{\rm cm^{-2}}.        \\
                \end{array}}
                \right.
\end{equation}

\section{Distributions of $RM_{\bar{p}\pi^+p}$ versus $M_{p\pi^0}$}

\begin{figure}[!htbp]
	\centering
	\subfigure[The distribution of $RM_{\bar{p}\pi^+p}$ versus $M_{p\pi^0}$ of the MC sample where $\Lambda$ from $\jpsi$ scatter inelastically with the nucleus in the target.]{ \includegraphics[width=0.45\textwidth]{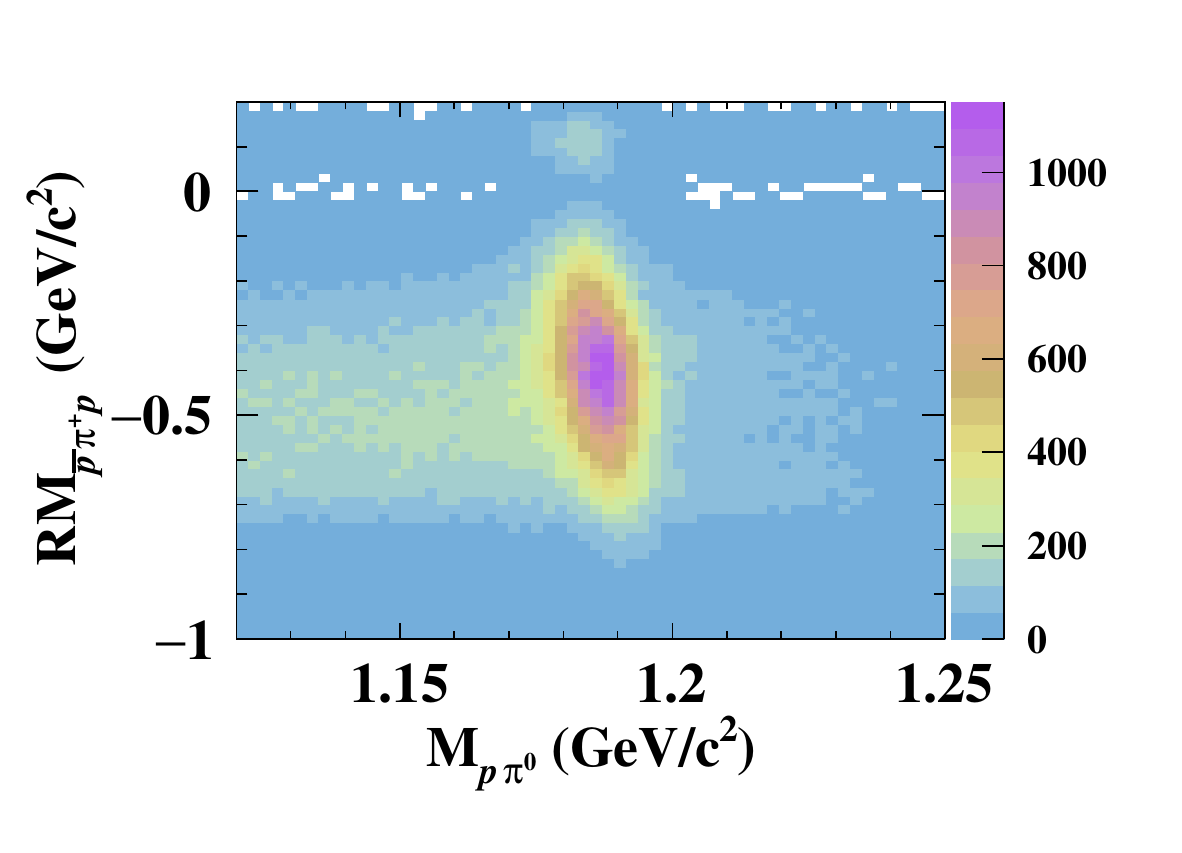} }
	\subfigure[The distribution of $RM_{\bar{p}\pi^+p}$ versus $M_{p\pi^0}$ of the MC sample where $\Lambda$ from $\jpsi$ decays into $p\pi^-$.]{ \includegraphics[width=0.45\textwidth]{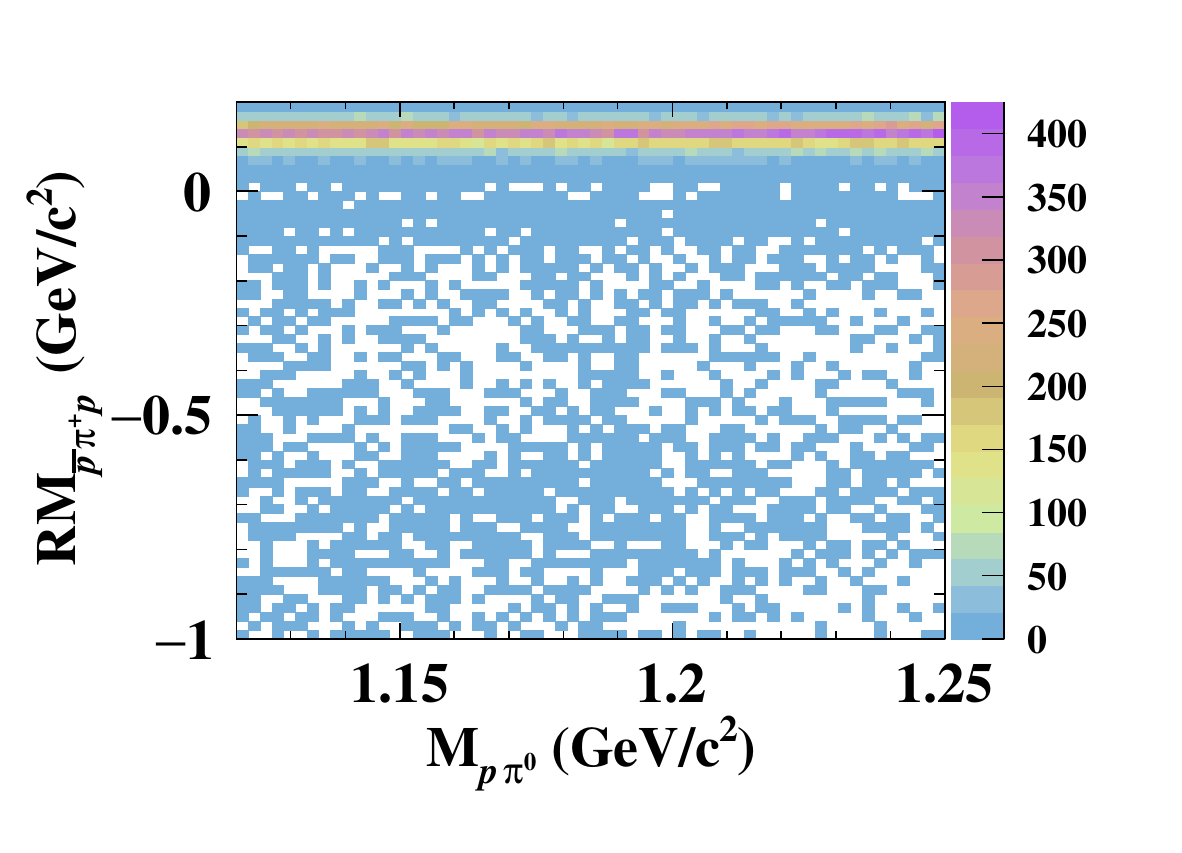} }
	\caption{}
	\label{fig:dis_mrecSgmP}
\end{figure}

The distributions of $RM_{\bar{p}\pi^+p}$ versus $M_{p\pi^0}$ are shown in Fig.~\ref{fig:dis_mrecSgmP}. If the incident $\Lambda$ does not scatter with nucleons, the recoil mass of $\bar{p} \pi^+ p$ should be around the known mass of $\pi^-$ according to the kinematic constraint. In inelastic scattering events we find that $RM_{\bar{p} \pi^+ p}$ will mostly be negative, which is explained by the additional mass and momentum contributed by nucleons in the reaction.

\clearpage


\begin{thebibliography}{**}

	\bibitem{Vidana:2018bdi} I.~Vida\~na,
\href{https://royalsocietypublishing.org/doi/10.1098/rspa.2018.0145}{Proc. Roy. Soc. Lond. A \textbf{474}, 0145 (2018)}.


	\bibitem{Hiyama:2018lgs} E.~Hiyama and K.~Nakazawa,
\href{https://www.annualreviews.org/doi/10.1146/annurev-nucl-101917-021108}{Ann. Rev. Nucl. Part. Sci. \textbf{68}, 131-159 (2018)}.


	\bibitem{Gal:2016boi} A.~Gal, E.~V.~Hungerford and D.~J.~Millener,
\href{https://journals.aps.org/rmp/abstract/10.1103/RevModPhys.88.035004}{Rev. Mod. Phys. \textbf{88}, no.3, 035004 (2016)}.


	\bibitem{Tolos:2020aln} L.~Tolos and L.~Fabbietti,
\href{https://www.sciencedirect.com/science/article/pii/S014664102030017X?via=ihub}{Prog. Part. Nucl. Phys. \textbf{112}, 103770 (2020)}.


	\bibitem{Eisele:1971mk} F.~Eisele, H.~Filthuth, W.~Foehlisch, V.~Hepp and G.~Zech,
\href{https://www.sciencedirect.com/science/article/pii/0370269371900530?via\%3Dihub}{Phys. Lett. B \textbf{37}, 204-206 (1971)}.


	\bibitem{Sechi-Zorn:1968mao} B.~Sechi-Zorn, B.~Kehoe, J.~Twitty and R.~A.~Burnstein,
\href{https://journals.aps.org/pr/abstract/10.1103/PhysRev.175.1735}{Phys. Rev. \textbf{175}, 1735-1740 (1968)}.


	\bibitem{Alexander:1968acu} G.~Alexander, U.~Karshon, A.~Shapira, G.~Yekutieli, R.~Engelmann, H.~Filthuth and W.~Lughofer,
\href{https://journals.aps.org/pr/abstract/10.1103/PhysRev.173.1452}{Phys. Rev. \textbf{173}, 1452-1460 (1968)}.


	\bibitem{Kadyk:1971tc} J.~A.~Kadyk, G.~Alexander, J.~H.~Chan, P.~Gaposchkin and G.~H.~Trilling,
\href{https://www.sciencedirect.com/science/article/pii/0550321371900769?via\%3Dihub}{Nucl. Phys. B \textbf{27}, 13-22 (1971)}.


	\bibitem{Hauptman:1977hr} J.~M.~Hauptman, J.~A.~Kadyk and G.~H.~Trilling,
\href{https://www.sciencedirect.com/science/article/pii/055032137790222X?via\%3Dihub}{Nucl. Phys. B \textbf{125}, 29-51 (1977)}.


	\bibitem{KEK-PSE-251:1997cno} J.~K.~Ahn \textit{et al.} [KEK-PS E-251],


	\bibitem{KEK-PS-E289:2000ytt} Y.~Kondo \textit{et al.} [KEK-PS-E289],
\href{https://www.sciencedirect.com/science/article/pii/S0375947400001913?via\%3Dihub}{Nucl. Phys. A \textbf{676}, 371-387 (2000)}.


	\bibitem{Ahn:2005jz} J.~K.~Ahn, S.~Aoki, K.~S.~Chung, M.~S.~Chung, H.~En'yo, T.~Fukuda, H.~Funahashi, Y.~Goto, A.~Higashi and M.~Ieiri, \textit{et al.}
\href{https://www.sciencedirect.com/science/article/pii/S0370269305018770?via\%3Dihub}{Phys. Lett. B \textbf{633}, 214-218 (2006)}.


	\bibitem{J-PARCE40:2021qxa} K.~Miwa \textit{et al.} [J-PARC E40],
\href{https://journals.aps.org/prc/abstract/10.1103/PhysRevC.104.045204}{Phys. Rev. C \textbf{104}, no.4, 045204 (2021)}.


	\bibitem{J-PARCE40:2021bgw} K.~Miwa \textit{et al.} [J-PARC E40],
\href{https://journals.aps.org/prl/abstract/10.1103/PhysRevLett.128.072501}{Phys. Rev. Lett. \textbf{128}, no.7, 072501 (2022)}.


	\bibitem{CLAS:2021gur} J.~Rowley \textit{et al.} [CLAS],
\href{https://journals.aps.org/prl/abstract/10.1103/PhysRevLett.127.272303}{Phys. Rev. Lett. \textbf{127}, no.27, 272303 (2021)}.


	\bibitem{J-PARCE40:2022nvq} T.~Nanamura \textit{et al.} [J-PARC E40],
[arXiv:\href{https://arxiv.org/abs/2203.08393}{2203.08393} [nucl-ex]].


	\bibitem{BESIII:2023clq} M.~Ablikim \textit{et al.} [BESIII],
\href{https://journals.aps.org/prl/abstract/10.1103/PhysRevLett.130.251902}{Phys. Rev. Lett. \textbf{130}, no.25, 251902 (2023)}

	\bibitem{ALICE:2022uso} S.~Acharya \textit{et al.} [ALICE],
\href{https://www.sciencedirect.com/science/article/pii/S0370269322003574?via\%3Dihub}{Phys. Lett. B \textbf{844} 137223 (2023)}


	\bibitem{Haidenbauer:2005zh} J.~Haidenbauer and Ulf-G.~Meissner,
\href{https://journals.aps.org/prc/abstract/10.1103/PhysRevC.72.044005}{Phys. Rev. C \textbf{72}, 044005 (2005)}.

	\bibitem{Rijken:2010zzb} T.~A.~Rijken, M.~M.~Nagels and Y.~Yamamoto,
\href{https://academic.oup.com/ptps/article/doi/10.1143/PTPS.185.14/1887789?login=false}{Prog. Theor. Phys. Suppl. \textbf{185}, 14-71 (2010)}.



	\bibitem{Polinder:2006zh} H.~Polinder, J.~Haidenbauer and U.~G.~Meissner,
\href{https://www.sciencedirect.com/science/article/pii/S0375947406006312?via\%3Dihub}{Nucl. Phys. A \textbf{779}, 244-266 (2006)}.


	\bibitem{Haidenbauer:2013oca} J.~Haidenbauer, S.~Petschauer, N.~Kaiser, U.~G.~Meissner, A.~Nogga and W.~Weise,
\href{https://www.sciencedirect.com/science/article/pii/S0375947413006167?via\%3Dihub}{Nucl. Phys. A \textbf{915}, 24-58 (2013)}.


	\bibitem{Haidenbauer:2015zqb} J.~Haidenbauer, U.~G.~Mei\ss{}ner and S.~Petschauer,
\href{https://www.sciencedirect.com/science/article/pii/S0375947416000075?via\%3Dihub}{Nucl. Phys. A \textbf{954}, 273-293 (2016)}.


	\bibitem{Haidenbauer:2018gvg} J.~Haidenbauer and U.~G.~Mei\ss{}ner,
\href{https://link.springer.com/article/10.1140/epja/i2019-12689-2}{Eur. Phys. J. A \textbf{55}, no.2, 23 (2019)}.


	\bibitem{Haidenbauer:2019boi} J.~Haidenbauer, U.~G.~Mei\ss{}ner and A.~Nogga,
\href{https://link.springer.com/article/10.1140/epja/s10050-020-00100-4}{Eur. Phys. J. A \textbf{56}, no.3, 91 (2020)}.


	\bibitem{Haidenbauer:2023qhf} J.~Haidenbauer, U.~G.~Mei\ss{}ner, A.~Nogga and H.~Le,
\href{https://link.springer.com/article/10.1140/epja/s10050-023-00960-6}{Eur. Phys. J. A \textbf{59}, no.3, 63 (2023)}.


	\bibitem{Li:2016paq} K.~W.~Li, X.~L.~Ren, L.~S.~Geng and B.~Long,
\href{https://journals.aps.org/prd/abstract/10.1103/PhysRevD.94.014029}{Phys. Rev. D \textbf{94}, no.1, 014029 (2016)}.


	\bibitem{Li:2016mln} K.~W.~Li, X.~L.~Ren, L.~S.~Geng and B.~W.~Long,
\href{https://iopscience.iop.org/article/10.1088/1674-1137/42/1/014105}{Chin. Phys. C \textbf{42}, no.1, 014105 (2018)}.



	\bibitem{Ishii:2006ec} N.~Ishii, S.~Aoki and T.~Hatsuda,
\href{https://journals.aps.org/prl/abstract/10.1103/PhysRevLett.99.022001}{Phys. Rev. Lett. \textbf{99}, 022001 (2007)}.


	\bibitem{Ishii:2012ssm} N.~Ishii \textit{et al.} [HAL QCD],
\href{https://www.sciencedirect.com/science/article/pii/S0370269312005266?via\%3Dihub}{Phys. Lett. B \textbf{712}, 437-441 (2012)}.



	\bibitem{Beane:2006gf} S.~R.~Beane \textit{et al.} [NPLQCD],
\href{https://www.sciencedirect.com/science/article/pii/S0375947407006434?via\%3Dihub}{Nucl. Phys. A \textbf{794}, 62-72 (2007)}.


	\bibitem{Beane:2010em} S.~R.~Beane, W.~Detmold, K.~Orginos and M.~J.~Savage,
\href{https://www.sciencedirect.com/science/article/pii/S0146641010000530?via\%3Dihub}{Prog. Part. Nucl. Phys. \textbf{66}, 1-40 (2011)}.



	\bibitem{Schaefer:2005fi} B.~J.~Schaefer, M.~Wagner, J.~Wambach, T.~T.~S.~Kuo and G.~E.~Brown,
\href{https://journals.aps.org/prc/abstract/10.1103/PhysRevC.73.011001}{Phys. Rev. C \textbf{73}, 011001 (2006)}.


	\bibitem{Fujiwara:2006yh} Y.~Fujiwara, Y.~Suzuki and C.~Nakamoto,
\href{https://www.sciencedirect.com/science/article/pii/S0146641006000718?via\%3Dihub}{Prog. Part. Nucl. Phys. \textbf{58}, 439-520 (2007)}.


	\bibitem{Lattimer:2000nx} J.~M.~Lattimer and M.~Prakash,
\href{https://iopscience.iop.org/article/10.1086/319702}{Astrophys. J. \textbf{550}, 426 (2001)}.


	\bibitem{Lonardoni:2014bwa} D.~Lonardoni, A.~Lovato, S.~Gandolfi and F.~Pederiva,
\href{https://journals.aps.org/prl/abstract/10.1103/PhysRevLett.114.092301}{Phys. Rev. Lett. \textbf{114}, no.9, 092301 (2015)}.


	\bibitem{LIGOScientific:2018cki} B.~P.~Abbott \textit{et al.} [LIGO Scientific and Virgo],
\href{https://journals.aps.org/prl/abstract/10.1103/PhysRevLett.121.161101}{Phys. Rev. Lett. \textbf{121}, no.16, 161101 (2018)}.


	\bibitem{Dai:2022wpg} J.~Dai, H.~B.~Li, H.~Miao and J.~Y.~Zhang,
[arXiv:\href{https://arxiv.org/abs/2209.12601}{2209.12601} [hep-ex]].


	\bibitem{Yuan:2021yks} C.~Z.~Yuan and M.~Karliner,
\href{https://journals.aps.org/prl/abstract/10.1103/PhysRevLett.127.012003}{Phys. Rev. Lett. \textbf{127}, no.1, 012003 (2021)}


	\bibitem{BESIII:2021cxx} M.~Ablikim \textit{et al.} [BESIII],
\href{https://iopscience.iop.org/article/10.1088/1674-1137/ac5c2e}{Chin. Phys. C \textbf{46}, no.7, 074001 (2022).}


	\bibitem{Li:2016tlt} H.~B.~Li,
\href{https://link.springer.com/article/10.1007/s11467-017-0691-9}{Front. Phys. (Beijing) \textbf{12}, no.5, 121301 (2017)}
[erratum: \href{https://link.springer.com/article/10.1007/s11467-019-0910-7}{Front. Phys. (Beijing) \textbf{14}, no.6, 64001 (2019)}].


	\bibitem{BESIII:2009fln} M.~Ablikim \textit{et al.} [BESIII],
\href{https://www.sciencedirect.com/science/article/pii/S0168900209023870?via\%3Dihub}{Nucl. Instrum. Meth. A \textbf{614}, 345-399 (2010)}

	\bibitem{supplemental} See Supplemental Material for additional information.


	\bibitem{geant4} S.~Agostinelli \textit{et al.} [{\sc GEANT4}],
\href{https://www.sciencedirect.com/science/article/pii/S0168900203013688?via\%3Dihub}{Nucl. Instrum. Meth. A \textbf{506}, 250 (2003).}

	\bibitem{detvis} K.~X.~Huang, {\it et al.},
\href{https://link.springer.com/article/10.1007/s41365-022-01133-8}{Nucl.\ Sci.\ Tech. \textbf{33}, 142 (2022)}.


	\bibitem{ref:kkmc} S.~Jadach, B.~F.~L.~Ward and Z.~Was,
\href{https://www.sciencedirect.com/science/article/pii/S0920563200008318?via\%3Dihub}{Nucl. Phys. B Proc. Suppl. \textbf{89}, 106 (2000)}.


	\bibitem{ref:evtgen} D.~J.~Lange,
\href{https://www.sciencedirect.com/science/article/pii/S0168900201000894}{Nucl. Instrum. Meth. A \textbf{462}, 152 (2001)}.


	\bibitem{ref:evtgen2} R.~G.~Ping,
\href{http://hepnp.ihep.ac.cn/article/doi/10.1088/1674-1137/32/8/001}{Chin. Phys. C \textbf{32}, 599 (2008)}.


	\bibitem{ParticleDataGroup:2022pth} R.~L.~Workman \textit{et al.} [Particle Data Group],
\href{https://academic.oup.com/ptep/article/2022/8/083C01/6651666?login=false}{PTEP \textbf{2022}, 083C01 (2022)}.


	\bibitem{ref:lundcharm} J.~C.~Chen, G.~S.~Huang, X.~R.~Qi, D.~H.~Zhang and Y.~S.~Zhu,
\href{https://journals.aps.org/prd/abstract/10.1103/PhysRevD.62.034003}{Phys. Rev. D \textbf{62}, 034003 (2000)}.


	\bibitem{ref:lundcharm2} R.~L.~Yang, R.~G.~Ping and H.~Chen,
\href{https://iopscience.iop.org/article/10.1088/0256-307X/31/6/061301}{Chin. Phys. Lett. \textbf{31}, 061301 (2014)}.


	\bibitem{BESIII:2022qax} M.~Ablikim \textit{et al.} [BESIII],
\href{https://journals.aps.org/prl/abstract/10.1103/PhysRevLett.129.131801}{Phys. Rev. Lett. \textbf{129}, no.13, 131801 (2022)}.


	\bibitem{Wright:2015xia} D.~H.~Wright and M.~H.~Kelsey,
\href{https://www.sciencedirect.com/science/article/abs/pii/S0168900215011134}{Nucl. Instrum. Meth. A \textbf{804}, 175-188 (2015)}.


%

	\bibitem{Millikan:1913zz} R.~A.~Millikan,
\href{https://journals.aps.org/pr/abstract/10.1103/PhysRev.2.109}{Phys. Rev. \textbf{2}, 109-143 (1913)}.

	\bibitem{NIST:2023} \href{https://webbook.nist.gov/chemistry/}{NIST Chemistry webbook}


	\bibitem{Barton:1982dg} D.~S.~Barton, G.~W.~Brandenburg, W.~Busza, T.~Dobrowolski, J.~I.~Friedman, C.~Halliwell, H.~W.~Kendall, T.~Lyons, B.~Nelson and L.~Rosenson, \textit{et al.}
\href{https://journals.aps.org/prd/abstract/10.1103/PhysRevD.27.2580}{Phys. Rev. D \textbf{27}, 2580 (1983)}.


	\bibitem{Cooper:1995ix} E.~D.~Cooper, B.~K.~Jennings and J.~Mares,
\href{https://www.sciencedirect.com/science/article/pii/0375947494005596?via\%3Dihub}{Nucl. Phys. A \textbf{585}, 157C-163C (1995)}.


	\bibitem{WA89:1997dmp} M.~I.~Adamovich \textit{et al.} [WA89],
\href{https://link.springer.com/article/10.1007/s002880050524}{Z. Phys. C \textbf{76}, 35-44 (1997)}.


	\bibitem{Botta:2001fu} E.~Botta for the OBELIX experiment,
\href{https://www.sciencedirect.com/science/article/abs/pii/S0375947401011575?via\%3Dihub}{Nucl. Phys. A \textbf{692}, 39-46 (2001)}.


	\bibitem{Astrua:2002zg} M.~Astrua, E.~Botta, T.~Bressani, D.~Calvo, C.~Casalegno, A.~Feliciello, A.~Filippi, S.~Marcello, M.~Agnello and F.~Iazzi,
\href{https://www.sciencedirect.com/science/article/abs/pii/S0375947401012520?via\%3Dihub}{Nucl. Phys. A \textbf{697}, 209-224 (2002)}.


	\bibitem{Lee:2018epd} T.~G.~Lee and C.~Y.~Wong,
\href{https://journals.aps.org/prc/abstract/10.1103/PhysRevC.97.054617}{Phys. Rev. C \textbf{97}, no.5, 054617 (2018)}.


\bibitem{BESIII:2021emv}
M.~Ablikim \textit{et al.} [BESIII],
\href{https://journals.aps.org/prd/abstract/10.1103/PhysRevD.104.072007}{Phys. Rev. D \textbf{104}, no.7, 072007 (2021)}.

\bibitem{BESIII:2010ank}
M.~Ablikim \textit{et al.} [BESIII],
\href{https://journals.aps.org/prd/abstract/10.1103/PhysRevD.81.052005}{Phys. Rev. D \textbf{81}, 052005 (2010)}.

\bibitem{Barlow:2002yb}
R.~Barlow,
[arXiv:\href{https://arxiv.org/abs/hep-ex/0207026}{hep-ex/0207026} [hep-ex]].

	\bibitem{Achasov:2023gey} M.~Achasov, \textit{et al.}
[arXiv:\href{https://arxiv.org/abs/2303.15790}{2303.15790} [hep-ex]].


%
%




\end{thebibliography}

\begin{thebibliography}{**}

\bibitem{CLAS:2021gur_2}
J.~Rowley \textit{et al.} [CLAS],
\href{https://journals.aps.org/prl/abstract/10.1103/PhysRevLett.127.272303}{Phys. Rev. Lett. \textbf{127}, no.27, 272303 (2021)}.

\bibitem{Barton:1982dg_2}
D.~S.~Barton, G.~W.~Brandenburg, W.~Busza, T.~Dobrowolski, J.~I.~Friedman, C.~Halliwell, H.~W.~Kendall, T.~Lyons, B.~Nelson and L.~Rosenson, \textit{et al.}
\href{https://journals.aps.org/prd/abstract/10.1103/PhysRevD.27.2580}{Phys. Rev. D \textbf{27}, 2580 (1983)}.

\bibitem{Cooper:1995ix_2}
E.~D.~Cooper, B.~K.~Jennings and J.~Mares,
\href{https://www.sciencedirect.com/science/article/pii/0375947494005596?via\%3Dihub}{Nucl. Phys. A \textbf{585}, 157C-163C (1995)}.

\bibitem{WA89:1997dmp_2}
M.~I.~Adamovich \textit{et al.} [WA89],
\href{https://link.springer.com/article/10.1007/s002880050524}{Z. Phys. C \textbf{76}, 35-44 (1997)}.

\bibitem{Botta:2001fu_2}
E.~Botta,
\href{https://www.sciencedirect.com/science/article/abs/pii/S0375947401011575?via\%3Dihub}{Nucl. Phys. A \textbf{692}, 39-46 (2001)}.

\bibitem{Astrua:2002zg_2}
M.~Astrua, E.~Botta, T.~Bressani, D.~Calvo, C.~Casalegno, A.~Feliciello, A.~Filippi, S.~Marcello, M.~Agnello and F.~Iazzi,
\href{https://www.sciencedirect.com/science/article/abs/pii/S0375947401012520?via\%3Dihub}{Nucl. Phys. A \textbf{697}, 209-224 (2002)}.

\bibitem{Lee:2018epd_2}
T.~G.~Lee and C.~Y.~Wong,
\href{https://journals.aps.org/prc/abstract/10.1103/PhysRevC.97.054617}{Phys. Rev. C \textbf{97}, no.5, 054617 (2018)}.



\end{thebibliography}
\end{document}